# Three New Fundamental Problems in Physics of the Aftershocks

Anatol V. Guglielmi, Alexey D. Zavyalov, Oleg D. Zotov

*Institute of Physics of the Earth RAS, Moscow, Russia*

[guglielmi@mail.ru](mailto:guglielmi@mail.ru), [zavyalov@ifz.ru](mailto:zavyalov@ifz.ru), [ozotov@inbox.ru](mailto:ozotov@inbox.ru)

## Abstract

Recently, the physics of aftershocks has been enriched by three new problems. We will conditionally call them dynamic, inverse, and morphological problems. They were clearly formulated, partially resolved and are fundamental. Dynamic problem is to search for the cumulative effect of a round-the-world seismic echo appearing after the mainshock of earthquake. According to the theory, a converging surface seismic wave excited by the main shock returns to the epicenter about 3 hours after the main shock and stimulates the excitation of a strong aftershock. Our observations confirm the theoretical expectation. The second problem is an adequate description of the averaged evolution of the aftershock flow. We introduced a new concept about the coefficient of deactivation of the earthquake source, "cooling down" after the main mainshock, and proposed an equation that describes the evolution of aftershocks. Based on the equation of evolution, we posed and solved the inverse problem of the physics of the earthquake source and created an Atlas of aftershocks demonstrating the diversity of variations of the deactivation coefficient. The third task is to model the spatial and spatio-temporal distribution of aftershocks. Its solution refines our understanding of the structure of the earthquake source. We also discuss in detail interesting unsolved problems in the physics of aftershocks.

# 1. Introduction

Numerous aftershocks accompanying strong earthquakes have been the subject of close attention since the birth of modern seismology [1, 2]. Suffice it to say that the first empirical law of earthquake physics was established by Fusakichi Omori, who studied aftershocks at the end of the century before last [3]. Nowadays, the interest in morphology and physics of aftershocks has intensified. And this is not surprising, since observation and analysis of aftershocks provide valuable information on the evolution of the source of a strong earthquake "cooling down" after a main shock.

In this paper, we outline three separate problems in the physics of aftershocks. We will conditionally call them dynamic, inverse, and morphological problem. They are presented in sections 2, 3 and 4 respectively. Each of them seems to us to be extremely interesting and unconditionally fundamental, and together they illuminate the properties of aftershocks from three different sides.

The dynamic problem relates to the field of investigation of triggers acting on the the earthquake source. It is customary to distinguish between endogenous and exogenous triggers [4]. To endogenous include, for example, seismic noise, always existing in the source after the main shock. Exogenous triggers are extremely diverse. These include seismic waves coming into a given earthquake source from other sources, as well as man-made and space impacts on the lithosphere [5–14]. (The literature on triggers is extensive. We have listed here for reference only a few publications.) The specific trigger which we will focus in this paper refers to a mixed type. On the one hand, it originates in the source after the main impact in the form of a surface seismic wave. On the other hand, it acts on the source only after some time, returning to the epicenter after propagation around the Earth [4, 10, 15–19]. Dynamic problem is to search for the cumulative effect of a round-the-world seismic echo appearing after the mainshock of earthquake. According to the theory, a converging surface seismic wave excited by the main shock returns to the epicenter about 3 hours after the main shock and stimulates the excitation of a strong aftershock. Our observations confirm the theoretical expectation.

The second problem is an adequate description of the averaged evolution of the aftershock flow. In 1894, the Omori law was formulated, according to which the frequency of aftershocks hyperbolically decreases over time [3]. After 30 years, Hirano proposed describing the evolution of aftershocks by a power function of the time elapsed after the main shock [20]. In the middle of the last century Utsu refined Hirano's methodology and successfully applied it to analyze the aftershocks [21, 22]. The history of discovery of the Omori law, as well as the modifications of the law in the works of Hirano and Utsu, are described in the papers [23–26].

In this paper we present a new approach to the problem.. We introduce a new concept about the coefficient of deactivation of the earthquake source, "cooling down" after the main shock, and propose an equation that describes the evolution of aftershocks [24, 27]. Based on the equation of evolution, we posed and solved the inverse problem of the physics of the earthquake source and created an Atlas of aftershocks demonstrating the diversity of variations of the deactivation coefficient [25, 28].

The third problem is to describe the structure and form of the spatial and spatio-temporal distribution of aftershocks. The amazing ribbed distribution structure, which was discovered through a careful analysis of the observations [29, 30], poses an interesting theoretical problem for seismology. In Section 5, we discuss possible approaches to solving this problem.

## 2. Seismic self-action of the earthquake source

In the Earth's lithosphere, there are many potential sources of earthquakes. Actualization of the particular potential source, i.e. the catastrophic formation of the rock continuity gap can occur spontaneously, but can also be induced by a seismic wave from another source in which such a gap has already occurred. The idea of induced transition of a source from a potential state to an actual underlies the concept of seismic interaction of earthquake sources. In all of this, we see a distant analogy with the idea of spontaneous and induced transitions in the atomic system [31], which is already interesting in itself, but can also serve as a model for seismicity modeling.

Interaction transfer is carried out, as a rule, by *diverging* seismic waves, so that the probability of the induced transition decreases with distance from the active source. There is one remarkable exception when the transfer is carried out by a *converging* surface wave. Such a wave is excited during the main shock of an earthquake. Having traveled around the Earth, the wave returns to the epicenter after about 3 hours. In this exceptional case, the wave that converges to the epicenter serves as a potent inducer of strong aftershocks because a circular converging wave steadily increases the energy density and the pulse approaches the epicenter. Thus, we are dealing with a nonlinear self-action of the earthquake source. The described process was called the *cumulative effect of a round-the-world seismic echo* [4, 16, 19].

The cumulative effect of round-the-world echo was observed by us after the corresponding processing of global catalogs, regional catalogs, and also by analyzing repeated shocks after individual strong earthquakes.

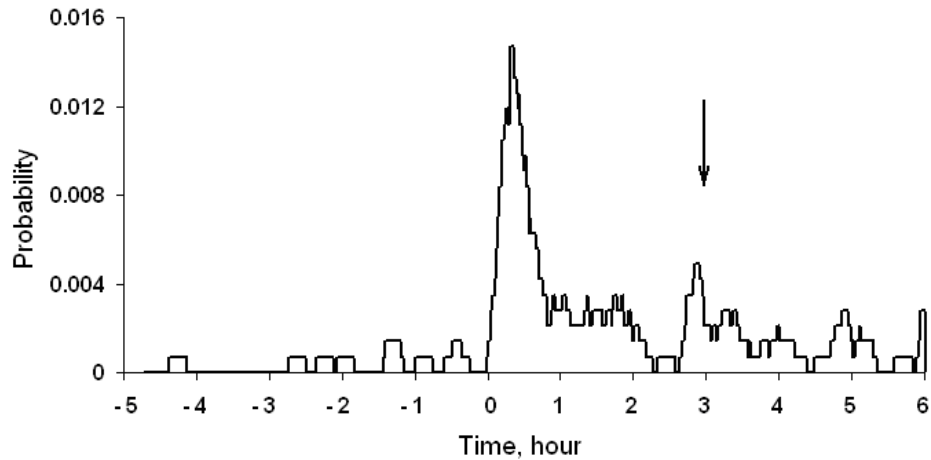

**Fig. 1.** The dynamics of the foreshocks and aftershocks with the magnitudes $6 \leq M < 7.5$ in the epicentral zones of 167 earthquakes with the magnitudes $M \geq 7.5$. The arrows mark the expected delay time of the around-the-world surface wave echo. The figure shows the probability density of the earthquake's occurrence as a function of time.

Figure 1 shows the probability density of the occurrence of aftershocks versus time over the interval of 6 hours after 167 earthquakes with magnitudes $M \geq 7.5$ according to the USGS catalog from 1973 to 2010. Here, we used the superposed epoch analysis. The occurrence times for the earthquakes with $M \geq 7.5$ are used as the references for synchronizing the aftershocks with magnitudes $6 \leq M < 7.5$ in the epicentral zones with a radius of 2°. We see that peak aftershock activity is observed during the first hour after a strong earthquake. This is followed by a certain attenuation and a new rise, which culminates in a new maximum in the aftershock activity about three hours later. This observation provides an argument in favor of our idea that the surface elastic waves excited by the main shock make a full rotation around the world and, having returned to the epicenter, can generate a strong aftershock there.

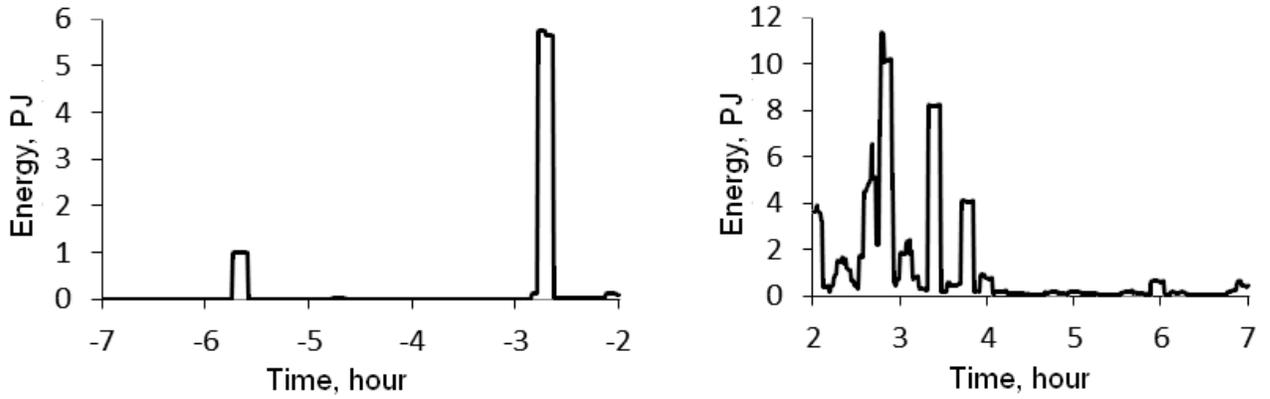

**Fig. 2.** The energy of the foreshocks (left panel) and aftershocks (right panel) in the epicentral zones of the strong earthquakes (M ≥ 7.5).

The notion of cumulative effect of the round-the-world seismic echo is not a pure abstraction. It is directly relevant to the reality [32]. Let us make more visual the empirical basis of this statement [33]. To do this, we transform the magnitude M into energy according to the known Gutenberg-Richter formula. The result is shown in Figure 2. The graph is obtained by the method of superimposed epoch analysis applied to the ISC catalog data for the earthquakes recorded from 1964 to 2009. The times of the earthquakes with magnitudes M≥ 7.5 are used as the time references for synchronizing the aftershocks and foreshocks. The figure displays the averaged time distribution of energy. The averaging was carried out by summing up the energy in 9-min intervals shifted along the time axis with a step of 1 min. By all appearances, in the right panel we see an intense cumulative effect of the converging surface waves excited by the main shock. The maximum energy reaches 11 PJ approximately 3 hours after the reference event (main shock). This corresponds to the idea of the effect of a round-the-world echo on the "cooling" source of an earthquake. The energy of the foreshocks is represented in the left panel of Figure 2 (see below).

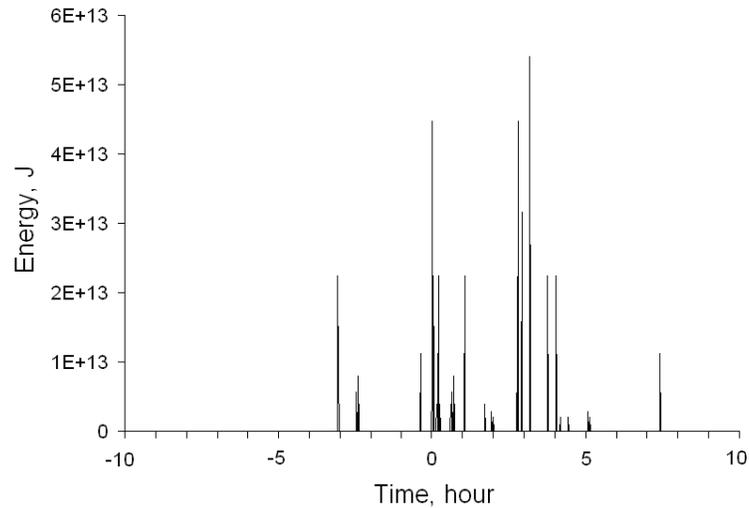

**Fig. 3.** The energy of foreshocks and aftershocks for the strong earthquakes in California. The zero hour corresponds to the instant of the main shock.

Something similar we also see in Figure 3. Here, in addition to the analyses of global seismicity, the figure also shows the results for the regional seismicity in California. We used catalog data presented at http://www.data.scec.org (1983–2008) and http://www.ncedc.org (1968–2007). The references to synchronize the foreshocks and aftershocks were specified by the time instants of the main events with magnitudes M ≥ 6 (zero hour in the figure). The earthquake energy was averaged in 20-min intervals shifted by 1-min steps. The energy of the main shocks, which attained a few petajoules in some events, was disregarded and not shown in the figure. About three hours after the reference, we see a high-power energy release by the aftershocks, which reaches a high of 54 TJ. This is also quite consistent with the idea of the influence of the round-the-world echo from the main shock on the "cooling" source of the earthquake.

We draw attention to a certain symmetry of the energy release of foreshocks and aftershocks relative to the moment of the main shock (see the Figures 2 and 3). Here we have in mind that the energy peaks observed for about 3 hours before the main shock and after 3 hours after it. This is a highly interesting. We know that the round-the-world echo of the main shock can stimulate the appearance of a strong aftershock at +3 h. Thus, it suggests that the echo signals from the foreshocks that form the peak at –3 h act as the triggers of the main shock, similar to the case when the echo-signals from the main shocks serve as the triggers for the aftershocks forming the peak at +3 h. It cannot be ruled out that at least parts of strong earthquakes are induced in this way.

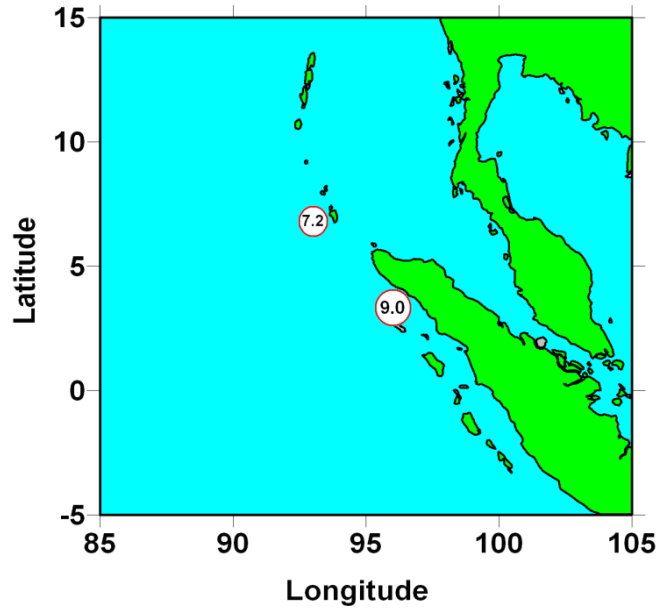

**Fig. 4.** The map of northeastern part of Indian Ocean with the epicenters of main shock of the Sumatra-Andaman earthquake (M = 9) and its strongest aftershock (M = 7.2).

We point out another observational fact in favor of our ideas [16]. Namely, we will demonstrate the cumulative effect of converging surface waves in the aftershock flow of the Sumatra-Andaman mega-earthquake – one of the strongest earthquakes of the beginning of the 21st century [34]. The earthquake with magnitude M = 9 occurred in Southeast Asia on December 26, 2004 at 00:58:53 UT. Its epicenter was located in the Indian Ocean, north of Simeulue Island, near the northwestern coast of Sumatra, as shown in Figure 4. The strongest aftershock (with the magnitude M = 7.2) was delayed by 3 h 20 min relative to the main shock (the epicenter of the aftershock is also shown in Figure 4). We noticed that the time delay of the aftershock is approximately equal to the travel time of the surface wave around the world. This suggested that the round-trip seismic echo signal could have been the trigger that initiated the strong aftershock.

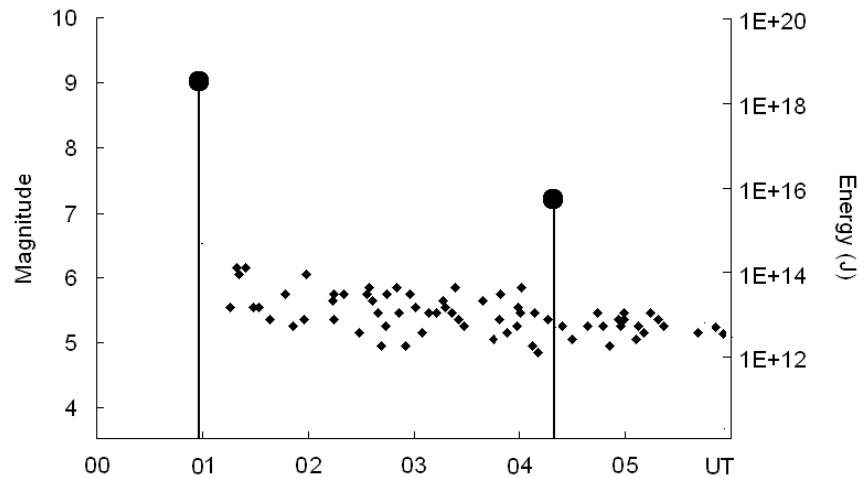

**Fig. 5.** The Sumatra-Andaman earthquake of December 26, 2004 and its aftershocks within a 5-h interval after the main shock. The black circles mark the main seismic shock and the strongest aftershock.

Figure 5 shows the aftershocks in the epicentral zone with a radius of 10°. The left vertical axis is scaled in units of magnitude, and the right vertical axis, in units of seismic energy estimated by the Gutenberg-Richter law. Seventy aftershocks were recorded within a 5-h interval. The strongest aftershock (M = 7.2) occurred 3 h 20 min after the main shock. It cannot be ruled out that this aftershock was induced by the round-the-world seismic echo. Our idea is that the surface elastic waves, which were excited by the main shock, have made a complete revolution around the globe, then returned to the vicinity of the epicenter and induced a strong aftershock there, whose occurrence was energetically prepared by the main shock.

## 3. Inverse problem

We introduce the parameter $\sigma$ characterizing the state of the earthquake source after the formation of the main rupture, and call it the s*ource deactivation coefficient* [24, 27]. The value of $\sigma$ indicates how fast the earthquake source loses its ability to excite aftershocks. We postulate the aftershock evolution law in the form of an integral relation

$$\int_0^t \sigma(t')dt' = g(t), \tag{1}$$

in which the auxiliary function $g(t)$ has the form

$$g(t) = [n_0 n(t)]^{-1}[n_0 - n(t)]. \tag{2}$$

Here, $n(t)$ is the frequency of aftershocks, $n_0 = n(0)$. The reference time is chosen, generally speaking, arbitrarily.

In (1) we have taken into account the possible time dependence of $\sigma$. The idea is to take into account phenomenologically the non-stationary of the geological environment during the relaxation of the source to a new equilibrium state. Nonstationarity may arise, in particular, under the influence of triggers.

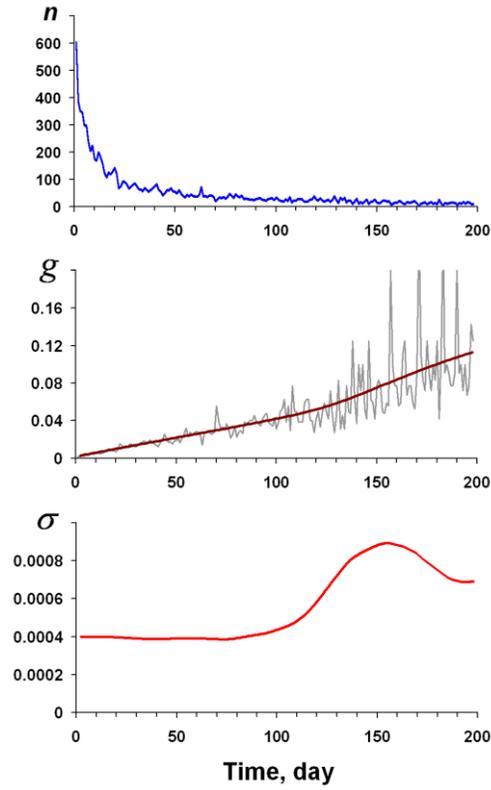

**Fig. 6.** An example of solving the inverse problem. The event took place in Southern California 17.01.1994. The magnitude of main shock is M = 6.7.

Considerations in favor of the proposed form (1) of the law of evolution will be given below. Before this, we look at (1) as the Volterra equation of the first kind with a trivial kernel. This immediately suggests us to pose the inverse problem of the physics of the source, i.e. try to determine the unknown function $\sigma(t)$ by the function $g(t)$ known from observations. Formally, the problem is solved trivially. However, we should take into account that our task, as often happens when solving inverse problems, is incorrectly posed. Regularization is smoothing auxiliary function $g(t)$. After regularization, the solution is

$$\sigma = \frac{d}{dt}\langle g \rangle, \qquad (3)$$

Here, the angle brackets denote smoothing. Figure 6 shows the example of solving the inverse problem [35].

However, let's return to the integral relation (1) and rewrite it in the form of a differential equation

$$\frac{dn}{dt} + \sigma n^2 = 0, \tag{4}$$

The general solution of equation (4) has the form

$$n(t) = n_0 \left[ 1 + n_0 \int_0^t \sigma(t') dt' \right]^{-1}, \tag{5}$$

It is easy to verify that (5) is a natural generalization of the Omori law. Indeed, if $\sigma = \text{const}$, then up to redefinitions formulas (1) and (5) coincide with the classical form of the aftershock evolution law

$$n(t) = \frac{k}{c + t}, \tag{6}$$

where $k > 0$, $c > 0$, $t \geq 0$ [3]. In contrast to (6), formulas (1) and (5) take into account the possible non-stationarity of rocks, which is expressed in the fact that the deactivation coefficient depends on time. We draw attention to the fact that formulas (1) and (5) preserve the hyperbolic structure of the law, only time flows unevenly.

We have outlined a project for creating an Atlas of aftershocks based on solving the inverse problem of the earthquake source [35]. The authors view the creation of the Atlas as a collective project, and call on interested seismologists to join the project. The processing of aftershocks by the proposed method is no more difficult than by the method of Omori or Hirano-Utsu (see, e.g., [20, 22, 23]). The analysis of information accumulated in the Aftershock Atlas will give interesting results. We believe that each sheet in this Atlas will contain information on the location, time of occurrence, and magnitude of a main shock, plots showing the variation for the coefficient of deactivation in relation to the time elapsed after the main shock, and other useful items. It is thought that an analysis of the information accumulated in the Atlas would add to a better understanding of the geodynamic processes in the rupture zone after the main shocks of earthquakes.

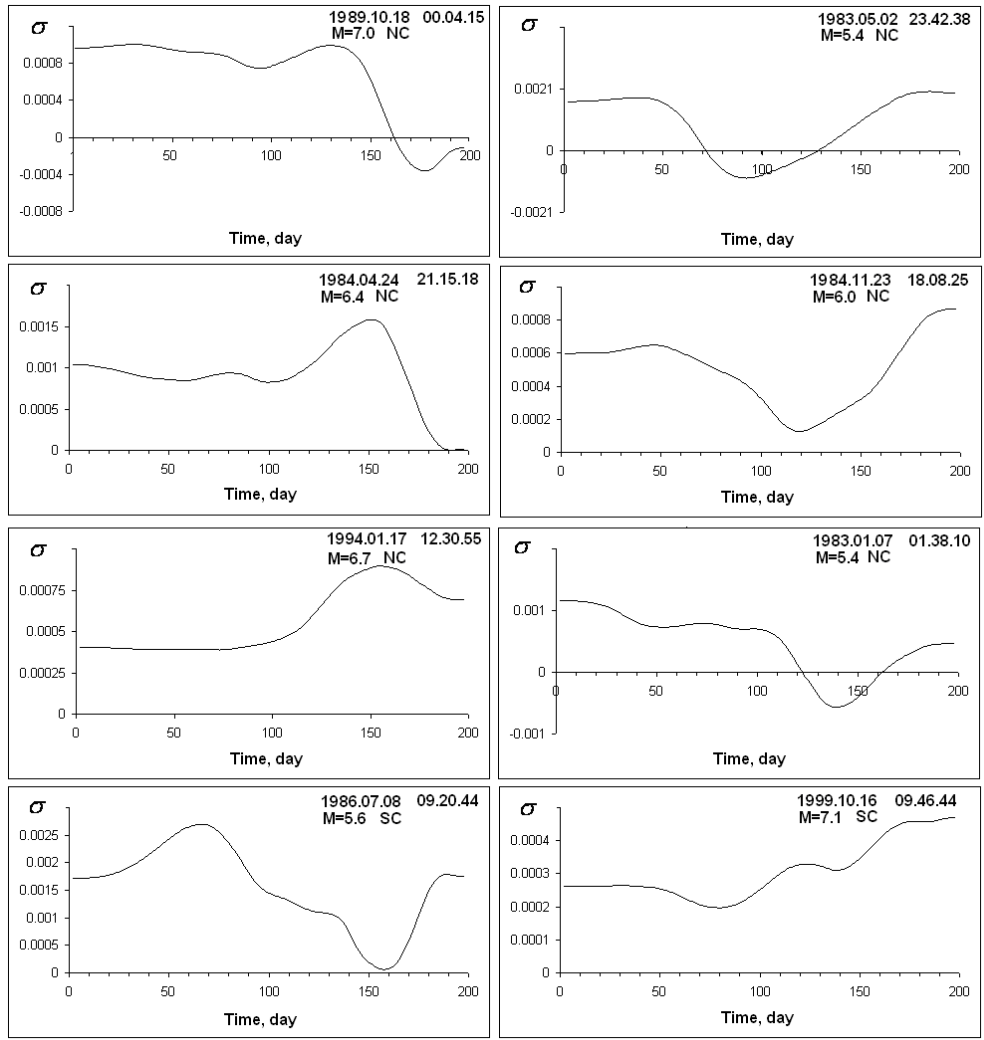

**Fig. 7.** The diversity in the forms of evolution for the coefficient of deactivation. The top right corner of each panel contains the year, month, and day of the main shock, the Greenwich time of occurrence, magnitude, and the catalog that was used (NC stands for northern California, and SC for southern California).

At present we have processed 20 events from the regional earthquake catalogs for northern California between 1968 and 2007 (http://www.ncedc.org) and for southern California for the period between 1983 and 2008 (https://www.scec.org). Figure 7 shows some fragments of the Aftershock Atlas. Here the deactivation coefficient $\sigma(t)$ can be seen for half of the processed events. This data selection shows that the deactivation coefficient experiences a complex evolution, occasionally going to negative values and then returning. We note that the negative values that are occasionally observed can be interpreted as periods of increased activity in the earthquake rupture zone.

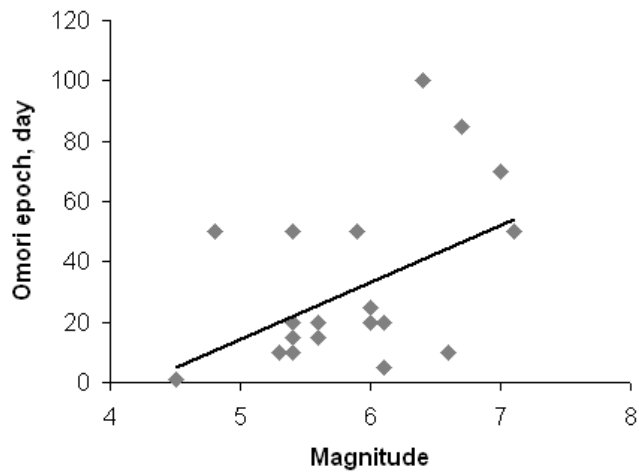

**Fig. 8.** Omori epoch duration versus magnitude of the main shock

` There are rather long Omori epochs ($\sigma = \text{const}$) in many cases. The median duration of the epoch is 25 days. The study found a weak tendency to increase the duration of the Omori epoch with increasing magnitude of the main shock (see the Figure 8). Summing up, we hope that the analysis of the information accumulated in the Atlas will yield interesting results that will shed light on the physics of the processes in the rupture zone of the main shock.

## 4. Spatial-temporal structure of aftershocks

The third problem, unlike the first two, has arisen unexpectedly. It falls out of the familiar context of seismology. Indeed, the idea of the self-action of the earthquake source originated on the basis of the well-known idea of the effect of one source on another, and the hypothesis about the possible dependence of the deactivation coefficient on time is naturally based on the Omori law. Moreover, the idea of the self-action of the source appeared along with the prediction of a strong aftershock 3 hours after the main shock. It was nothing like this, when we are suddenly confronted with the third problem. It was not preceded by any physical considerations. The problem arose accidentally during a morphological study of the spatial distribution of aftershocks averaged over time [29, 30].

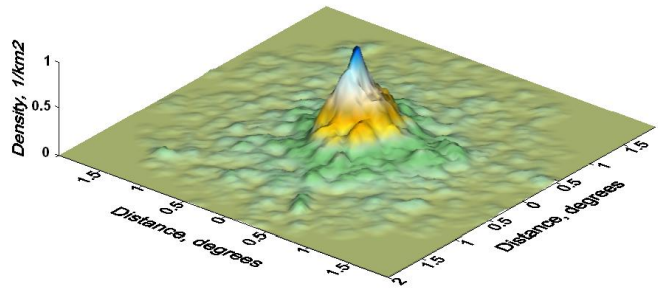

**Fig. 9.** Generalized spatial distribution of aftershocks according to the USGS catalog (1974-2014).

Figure 9 shows a generalized picture of the spatial distribution of aftershocks in the vicinity of the epicenter of the main shock. The horizontal plane coincides with the earth's surface, and the center of the coordinate system coincides with the epicenter. The normalized surface density of aftershocks is plotted along the vertical axis. USGS catalog data for the period from 1973 to 2014 was used to construct the density distribution. 190 main shocks with a magnitude of $M \geq 7.5$ and about 6000 aftershocks in a circle with a radius of 2 degrees (about 200 km) were selected over a time interval of 10 hours after the main shocks. The generalized spatial distribution of aftershocks is obtained by the epochs superposition method. The coordinates of the main shocks were used as a reference point.

We see that the generalized distribution has a domed shape. It was found that the effective width of the distribution L depends on the magnitude of the main shock M as follows:

$$\lg L[\text{km}] = 0.43M - 1.27,  \qquad (7)$$

The relation (7) can be used to determine the characteristic size of the focal zone, at least in the magnitude range M ≥ 5.5 [29]. Our relationship is consistent with known dependencies of this kind (see, for example, [36, 37]).

So, we have a general idea about the evolution of aftershocks averaged over the space, and about the spatial distribution averaged over the time. An attempt to present together the spatial and temporal distributions of strong aftershocks in one drawing led to an unexpected result. It is shown in Figure 10.

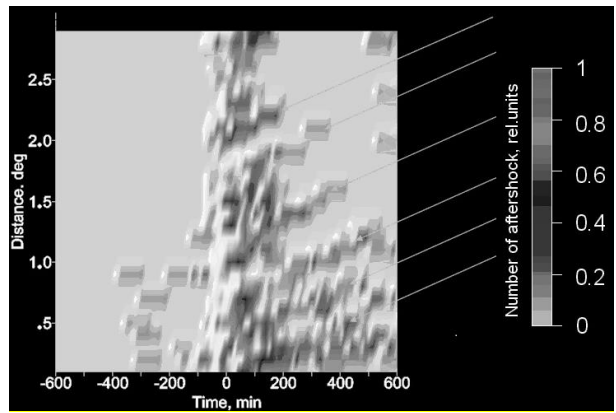

**Fig. 10.** The distribution of strong aftershocks in the time-distance coordinates. Distance is given in degrees of the large circle arc. Thin white lines indicate the observed structure.

Figure 10 shows the spatial-temporal distribution of the 208 aftershocks with 6 ≤ Maft < Mms after 190 main shocks with M ≥ 7.5 in the time-distance coordinates. Distance is given in degrees of the large circle arc. We see that in the spatio-temporal evolution of strong aftershocks, a similarity of the wave structure is observed. The spatial period of the ribbed structure is several tens of kilometers. The velocity of propagation along structural elements is approximately 7–10 km / h.

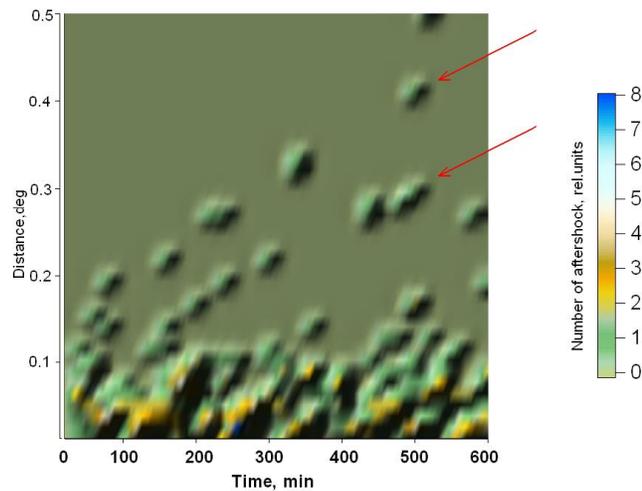

**Fig. 11.** Spatial-temporal structure of aftershocks после главного толчка M=6.6. in Southern California. Red arrows indicate the observed structural elements.

Figure 10 was built according to the world catalog of earthquakes. It is important to note that the unusual structure of the spatiotemporal distribution of aftershocks is also observed in the study of regional catalogs. As an example, we give here Figure 11, which shows the spatiotemporal structure of aftershocks for the earthquake with magnitude M = 6.6, which occurred in Southern California on 1987.11.24.

Summarizing the foregoing, we can say that the third problem consists primarily in an independent verification of the adequacy of our description of the structure and form of the spatio-temporal distribution of aftershocks. The unusual ribbed distribution structure, which apparently was discovered through a careful analysis of the observations, poses an interesting theoretical problem for seismology.

## 5. Discussion

The phenomena described above were discovered as a result of the experimental study. In this section of the paper we discuss each of them. But before that, we want to emphasize the reproducibility of the experiment and the repeatability of the results. These two essential properties of experimental work inspire us with confidence that we are dealing with real phenomena. On the other hand, we are fully aware that our results should be independently verified.

### 5.1. Echo

The phenomenon of round-the-world echo in the form of a strong aftershock is certainly significant from a geophysical point of view. The parameters of a circular surface wave converging to the epicenter are quite amenable to control. Therefore, the echo can be used as a test signal to study the nonlinear response of the earthquake source to impulse action.

Perhaps the phenomenon we discovered has some practical significance. It is known that the production of rescue operations after a catastrophic earthquake is accompanied by the threat of the collapse of dilapidated buildings as a result of afterdhocks. We predict the likely occurrence of a strong aftershock approximately 3 hours after the main shock. Therefore, it is advisable to stop rescue operations from 2 hours 40 minutes to 3 hours 20 minutes after the main shock in order to reduce the risk of death of rescuers.

In connection with the phenomenon of the self-action of the source by means of a round-the-world echo, it is appropriate to raise the question of the possible existence of relaxation auto-oscillations of the Earth. We notice that the very fact that near-hourly free oscillations $_0S_2$ affect global seismicity does not necessarily mean that *resonant* self-excited oscillations of the Earth do exist [15, 16]. However, the question of the existence of three-hour *relaxation* self-oscillations is much less clear [17].

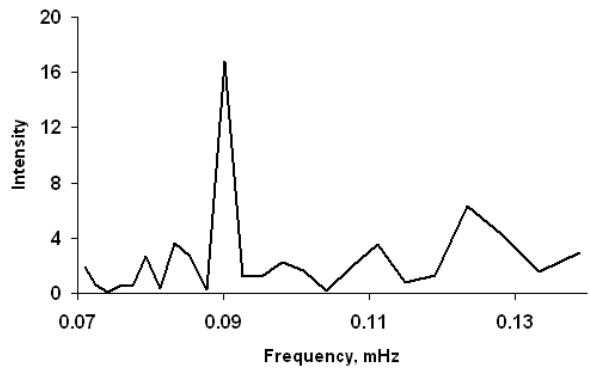

**Fig. 12.** The spectrum of global seismicity according to the USGS catalog (1973-2014). The figure shows the frequency dependence of the intensity of the Fourier components $|n_\omega|^2$. Processed 18835 events with magnitudes $M \geq 5.5$.

Multiple energy release peaks with a quasi-period of 3 hours were observed by us after two strong earthquakes in Sumatra and Japan [16]. A similar effect was found statistically in California. Moreover, a peak at a frequency of 0.09 mHz is clearly present in the global seismicity spectrum, which corresponds to a period of 3 hours (see Figure 12). These facts plausibly suggest that a quasi-period of 3 h on average is a new dynamically important characteristic of the Earth as an oscillating body, previously not known.

It is plausible that the aftershock induced by the round-the-world echo excites a secondary surface wave. Constructive interference of the primary wave and secondary wave enhances the cumulative effect. In other words, regeneration of the round-the-world echo takes place.

We cannot state for sure that this would certainly result in the formation of the self-excited relaxation oscillator, which works by the scheme described above and outputs the pulsed signals with a period of 3 h. At the same time, the idea seems to be promising and, in our opinion, deserves closer consideration.

The question is whether the energy of the echo-induced aftershocks is sufficient to fully compensate the energy losses of the surface wave, which have made a complete circuit around the Earth? If these losses are compensated only partially, the epicentral zone of the lithosphere would act as a relaxation oscillator in the underexcited state. It is known from radiophysics that the underexcited oscillator can act as a regenerative (positive-feedback) amplifier of the oscillations; however, we only note this in order to give an impetus to further discussion.

Finally, we briefly discuss the cumulative effect from the standpoint of geomagnetism. Previously, excitation of ULF magnetic pulsations by seismic waves *divergent* from the earthquake source was studied in the context of seismoelectrodynamics [38]. Quite clearly, the experimental identification and theoretical modeling of the strong pulses of the ULF magnetic field excited by the collapse of the *convergent waves* is of undoubted interest.

### 5.2. Relaxation theory of deactivation

The aftershock equation (4) includes a phenomenological parameter, namely, the deactivation coefficient $\sigma$. It would be desirable to interpret $\sigma$ in the framework of geodynamics and tectonophysics, but this seems to be an extremely difficult task. If we remain at the phenomenological level, then we will have the opportunity to propose a model for the evolution of the deactivation coefficient.

Let us use the distant analogy between the variations of the state of rocks in the earthquake source and the variations of the Earth's climate [39]. Suppose that there is an equilibrium state $\bar{\sigma}$, generally speaking, depending on time. Let $\tau$ is the characteristic time approached $\sigma$ to equilibrium. Then the relaxation theory of deactivation can be based on the equation

$$\frac{d\sigma}{dt} = \frac{\bar{\sigma}(t) - \sigma}{\tau} + \xi(t), \tag{8}$$

similar to that used in climatology to describe the average temperature on the earth's surface. Function $\xi(t)$ simulates the effect on the earthquake source of endogenous and exogenous triggers.

In its time, it has come to be believed in geophysics that two identical geomagnetic storms are barely possible. Each storm has its own unique combination of a huge number of local and global parameters of the near-Earth environment. Perhaps, this is also true for strong earthquakes, since the nucleation and activation of the source depend on a very broad variety of the physical conditions in the seismogenerating structures of the Earth's crust [40]. In light of this, it appears important to study the deactivation coefficient based on the empirical data as a function of the tectonic position and geological structure of the source.

### 5.3. Ribbed structure

We turn to the discussion of the third problem. The amazing morphology of the spatiotemporal distribution that we discovered undoubtedly requires a reasonable interpretation. But in this case, we do not have a phenomenological theory within which we could understand the formation of the ribbed distribution structure at least at a qualitative level. Further work remains to be done both on refining the morphology of the distribution and on the physical understanding of the result.

As a phenomenological palliative, we propose to consider the equation of nonlinear diffusion [41, 42]. Let us suppose that a diffusion process that is not yet known to us forms a spatial distribution. Our completely trivial idea is to add the diffusion term to the right side of the evolution equation (4):

$$\partial n / \partial t = -\sigma n^2 + \hat{D}\nabla^2 n, \qquad (9)$$

$$\hat{D} = \begin{Vmatrix} D_{\parallel} & 0 \\ 0 & D_{\perp} \end{Vmatrix}, \qquad (10)$$

where $\nabla$ is 2D Hamilton operator. Here we have taken into account the anisotropy of faults in the earth crust at the phenomenological level [24, 30]. The nonlinear diffusion equation (9) has a rich set of solutions, but so far it is difficult to say whether at least one of them is suitable for modeling the distributions shown in Figures 10 and 11. Thus, equation (9) gives us only a temporary improvement in the state of the third problem.

The unsolved problems of the physics of aftershocks are not limited to those that we have indicated above. It would be interesting to discuss the impact of anthropogenic [5–7, 9–13] and space factors [8, 13, 14] on the activity of aftershocks, but this would lead us far away from the topic of this paper.

## 6. Conclusion

We presented three new problems in the physics of aftershocks. Each of them is interesting in itself, and all together they expand the conceptual baggage of seismology. The first two problems are partially solved by fairly simple means. However, despite the apparent simplicity, they deserve close attention, since they provide an opportunity to take a fresh look on the most complex physical processes that occur in the earthquake source after the formation of the gap of rock continuity during the main shock.

The third problem stands apart. Detection of the ribbed structure of the spatiotemporal distribution aftershocks yet not amenable to rational explanation. We are faced with a mystery that has yet to be solved by joint efforts. The palliative solution proposed by us may turn out to be erroneous, since it is based on an arbitrary assumption.

*Acknowledgments*. We express our deep gratitude to A.L. Buchachenko, B.I. Klain and A.S. Potapov for numerous fruitful discussions. The work was supported by the projects of the RFBR 18-05-00096 and 19-05-00574, Program No. 12 of the RAS Presidium, as well as the state assignment program of the IPhE RAS.